\documentclass[11pt, a4paper]{article}

\usepackage[margin=1in]{geometry}
\usepackage{tabularx}
\newcolumntype{Y}{>{\centering\arraybackslash}X}
\usepackage{graphicx}
\usepackage{booktabs}
\usepackage{multirow}
\usepackage[table]{xcolor}
\usepackage{xcolor}
\usepackage{float}
\usepackage{amsmath,amssymb,amsfonts,amsthm}
\usepackage{mathrsfs}
\usepackage{changes}
\usepackage{textcomp}
\usepackage[version=4]{mhchem}
\usepackage[T1]{fontenc}
\usepackage[utf8]{inputenc}
\usepackage{authblk}
\usepackage{fontawesome5}
\usepackage[colorlinks=true,linkcolor=blue,citecolor=blue,urlcolor=blue]{hyperref}

\usepackage{algorithm}
\usepackage{algorithmicx}
\usepackage{algpseudocode}
\usepackage{listings}

\begin{document}

\bibliographystyle{unsrt} 

\title{Benchmarking Universal Machine-Learned Interatomic Potentials for High-Temperature Metal-Organic Framework Chemistry}

\author[1]{Connor W. Edwards}
\author[1]{Jack D. Evans\thanks{Corresponding author: j.evans@adelaide.edu.au}}

\affil[1]{School of Physics, Chemistry and Earth Sciences, Adelaide University, North Terrace, Adelaide, 5000, South Australia, Australia}

\date{}

\maketitle

\abstract{
Universal machine-learned interatomic potentials (uMLIPs) offer a promising approach to performing atomistic simulations at near-DFT accuracy with greatly reduced computational cost. Here, we present a new high-temperature benchmarking dataset of 40~ps ab~initio molecular dynamics (AIMD) trajectories simulated at 300, 1000, and 2000~K for nine zinc- and zirconium-based metal-organic frameworks (MOFs): ZIF-8, CALF-20, MOF-10, MOF-5, MIP-206, UiO-66, UiO-67, UiO-66-\ce{NH2}, and NU-1000. These trajectories capture equilibrium dynamics, thermally induced distortions, and early-stage decomposition events, including linker degradation and metal node aggregation. Subsequently, we use this dataset to benchmark five leading uMLIPs: ORB-v3, MACE-MP-0a, MACE-MPA-0, fairchem ODAC23, and fairchem OMAT. Our results reveal that ORB-v3 and fairchem OMAT achieve the lowest energy, force, and stress errors across all temperatures. However, all models exhibit significant error under high-temperature conditions. Long-timescale molecular dynamics simulations produced with ORB-v3 demonstrate that the generative error of uMLIPs far exceeds model losses captured during static validation, highlighting the limitations of current universal models for simulating high-temperature MOF dynamics. This work provides a benchmark for assessing the robustness of uMLIPs in extreme regimes and guides future development of potentials capable of accurately modeling the chemistry of high-temperature MOF dynamics.
}
\newpage

\section{Introduction}
Understanding how metal–organic frameworks (MOFs) decompose at high temperatures is critical for designing MOF-derived catalysts, yet the atomic-scale details of this process have remained largely inaccessible to experiment. MOFs are porous crystalline solids whose structure and chemistry can be tuned by varying their metal nodes and organic linkers,\cite{10.1016/j.ccr.2020.213319, 10.1021/ar5000314, 10.1126/science.1230444, 10.1038/nature01650} a feature that has driven their exploration in catalysis.\cite{Bavykina2020} When subjected to pyrolysis or calcination, MOFs transform into amorphous materials that frequently surpass their crystalline parents in catalytic performance.\cite{Konnerth2020, Li2021, 10.1021/acsami.5c10085, 10.1021/acs.chemrev.9b00685} The disordered nature of these products poses a fundamental characterization challenge. For example, conventional diffraction and spectroscopic methods provide limited insight into amorphous structures.\cite{10.1021/acs.jpcc.2c01091, 10.1039/C7CP08508G, 10.1021/jacs.9b03234} Although metal nanoparticles, reduced organics, and carbon networks have been identified as decomposition products,\cite{10.1021/acsami.5c10085, 10.1021/acscatal.6b01222} the transformation mechanisms that govern their formation require further elucidation.

Molecular simulation offers a method to directly probe these processes on the atomistic scale. Capturing MOF pyrolysis demands access to large spatial and temporal scales, necessitating a method that is both accurate and efficient to describe the high energy, non-equilibrium states. While density functional theory (DFT) is accurate, it is computationally expensive and typically limited to simulations of smaller systems on the order of tens of picoseconds.\cite{10.1126/science.abn3445, 10.26434/chemrxiv-2024-n2vzq, 10.1021/acs.jpcc.2c01091} This large computational cost renders DFT impractical for observing complete MOF pyrolysis. However, even on short timescales ab~initio molecular dynamics (AIMD) has been able to capture the early stages of decomposition pathways, including gas formation and zinc environments in ZIF-8 and CALF-20, at 2000~K.\cite{10.1002/adts.202500514}

Machine-learned interatomic potentials (MLIPs) provide a computationally efficient alternative, enabling molecular dynamics simulations at near-DFT accuracy over nanosecond timescales.\cite{10.1039/d3dd00236e, 10.1021/acs.jpclett.4c00746, 10.1038/s41563-020-0777-6} Training such models from scratch, however, requires large volumes of high-fidelity reference data, which is computationally demanding to generate. This has motivated the development of universal MLIPs (uMLIPs) trained on large, diverse datasets such as MPtrj and sAlex, which contain millions of inorganic and molecular structures.\cite{salex_2024, mptrj_2023} These uMLIPs function as versatile ``off-the-shelf'' tools for atomistic simulations. Several prominent uMLIP families have emerged recently, including the MACE, ORB, and fairchem UMA model suites.\cite{10.48550/arXiv.2205.06643, Batatia2022mace, 10.48550/ARXIV.2504.06231, 10.5281/zenodo.15587498, 10.1021/jacs.4c14455}
Model performance is constantly tested by the community and one benchmark, MATBench, currently ranks them from highest to lowest accuracy: fairchem, ORB-v3, MACE-MPA-0 and MACE-MP-0a.\cite{10.48550/arXiv.2308.14920} 

These uMLIPs have demonstrated impressive transferability across materials and in the prediction of structural properties.\cite{Kra2025, Loew2025, 10.48550/arxiv.2511.22885} However, current uMLIPs face significant challenges when applied to high-temperature regimes.\cite{10.1002/adts.202500514, arxiv.2601.16459} The vast majority of available training datasets consist of near equilibrium, low energy structures sampled from relaxation trajectories.\cite{mptrj_2023} High energy regions of the potential energy surface including bond-breaking events and intermediates or amorphous structures are sparsely represented or entirely absent, leading to a softening of the potential energy surface (PES).\cite{Deng2025} Consequently, universal models are unable to treat high-temperature pathways at DFT level accuracy.\cite{10.1002/adts.202500514} This leads to unstable molecular dynamics trajectories, inaccurate predictions of energies and interatomic forces, and unphysical predictions of structural features.\cite{arxiv.2601.16459} Newer datasets such as OMAT24 and ODAC25 extend previous datasets to include non equilibrium structures and systems with multiple gaseous adsorbates.\cite{10.48550/arxiv.2508.03162, 10.48550/arXiv.2410.12771} While universal models trained on these datasets show improvement over models trained on MPtrj, they are still unable to recreate high temperature DFT calculations or \ce{CO2} isotherms.\cite{arxiv.2601.16459, 10.48550/arxiv.2602.13725} However, no systematic high-temperature DFT dataset currently exists for benchmarking the performance of MLIPs under thermally driven MOF transformations. Addressing this gap requires datasets that capture realistic high temperature chemistry while providing sufficient volumes of accurate DFT-level data for rigorous validation.

In this work, we introduce a collection of high temperature AIMD trajectories designed for benchmarking uMLIPs. This dataset comprises 27 AIMD trajectories spanning nine chemically diverse zinc and zirconium MOFs; ZIF-8, CALF-20, MOF-10, MOF-5, MIP-206, UiO-66, UiO-67, UiO-66-\ce{NH2} and NU-1000. Each MOF was simulated at 300, 1000, and 2000~K to capture equilibrium dynamics, thermally activated distortions, and early stage decomposition pathways. This represents one of the first systematic efforts to generate large scale, high fidelity trajectories for MOFs under extreme temperature conditions. Using this data, we evaluate leading uMLIPs, characterizing their accuracy, dynamical stability, and behavior across this challenging out-of-domain regime. Our results establish the current limits of universal atomistic modeling for high-temperatures and provide guidance for improving the robustness and applicability of uMLIPs for this application and more.

\section{Computational methods}
To investigate the dynamics of ZIF-8, CALF-20, MOF-10, MOF-5, MIP-206, UiO-66, UiO-67, UiO-66-\ce{NH2} and NU-1000, a series of high temperature trajectories were created using the quickstep module of CP2K, version 2023.2.\cite{10.1016/j.cpc.2004.12.014, 10.1063/5.0007045} Reference geometry and cell optimizations were performed on all MOFs. Valence electrons were described using the triple-$\zeta$ valence polarized basis sets and norm-conserving Goedecker-Teter-Hutter pseudopotentials.\cite{10.1103/PhysRevB.54.1703} The PBE approximation was used for evaluating exchange correlation energy and pair potential interactions were treated by DFT-D3.\cite{10.1103/PhysRevLett.77.3865, 10.1063/1.3382344} The optimization was computed using the Broyden-Fletcher-Goldfarb-Shanno (BFGS) optimiser with a 1000 Ry plane wave cutoff, a 60 Ry relative cutoff and an SCF convergence criterion of $10^{-6}$~Hartree. The convergence criteria for optimization were a maximum force convergence of $4.5 \times 10^{-5}$~Hartree$\,$Bohr$^{-1}$ and a root mean square convergence of $3 \times 10^{-5}$~Hartree$\,$Bohr$^{-1}$. From these optimized structures, molecular dynamics trajectories in the NVT ensemble were generated at 300~K, 1000~K and 2000~K for each MOF. We used a 0.5~fs timestep and the canonical sampling through velocity rescaling (CSVR) thermostat with a time constant of 1000~fs for a total of 40~ps ({80,000}~steps).\cite{10.1063/1.2408420} For the dynamics, the plane wave cutoff was updated to 500~Ry and the relative cutoff was updated to 50~Ry.

All calculations in this study were performed using MACE-torch (version 0.3.14),\cite{10.48550/arXiv.2401.00096} orb-models (version 0.5.5),\cite{10.48550/arXiv.2410.22570} and fairchem (version 2.13.0).\cite{10.5281/zenodo.15587498} All calculations and simulation trajectories were managed using the Atomic Simulation Environment (ASE) version 3.26.0.\cite{ase-paper} These models were validated by sampling a random 1000 structures from each MOF-temperature AIMD dataset. 

Ramp simulations of 1~ns  were performed using ORB-v3 for all nine MOFs with ASE employing the Langevin thermostat with a 0.5~fs timestep and a friction coefficient of 0.01~fs$^{-1}$.  These simulations involved three regimes; 300~K for 300~ps, ramping from 300 to 2000~K for 200~ps at a rate of 8.5~K$\,$ps$^{-1}$ and then running at 2000~K for 500~ps. Each regime was divided into 100 equal time intervals, and one structure was randomly sampled from each interval. Reference energy force and stress values were calculated using the same DFT methods as for the AIMD simulations. Where loss is calculated, the energy:force:stress weights used are 1:100:1 with energy in eV, force in eV~\AA$^{-1}$ and stress in eV~\AA$^{-3}$.

We previously reported 16~ps AIMD trajectories for ZIF-8 and CALF-20.\cite{10.1002/adts.202500514} The present work extends those simulations to 40~ps and expands coverage to seven additional MOF systems, providing substantially longer trajectories that capture later-stage decomposition events not accessible in the earlier dataset.

\section{Results and discussion}

\subsection{Characterizing the trajectories}

\begin{figure}[tbp]
    \vspace{5mm}
    \centering
    \includegraphics[]{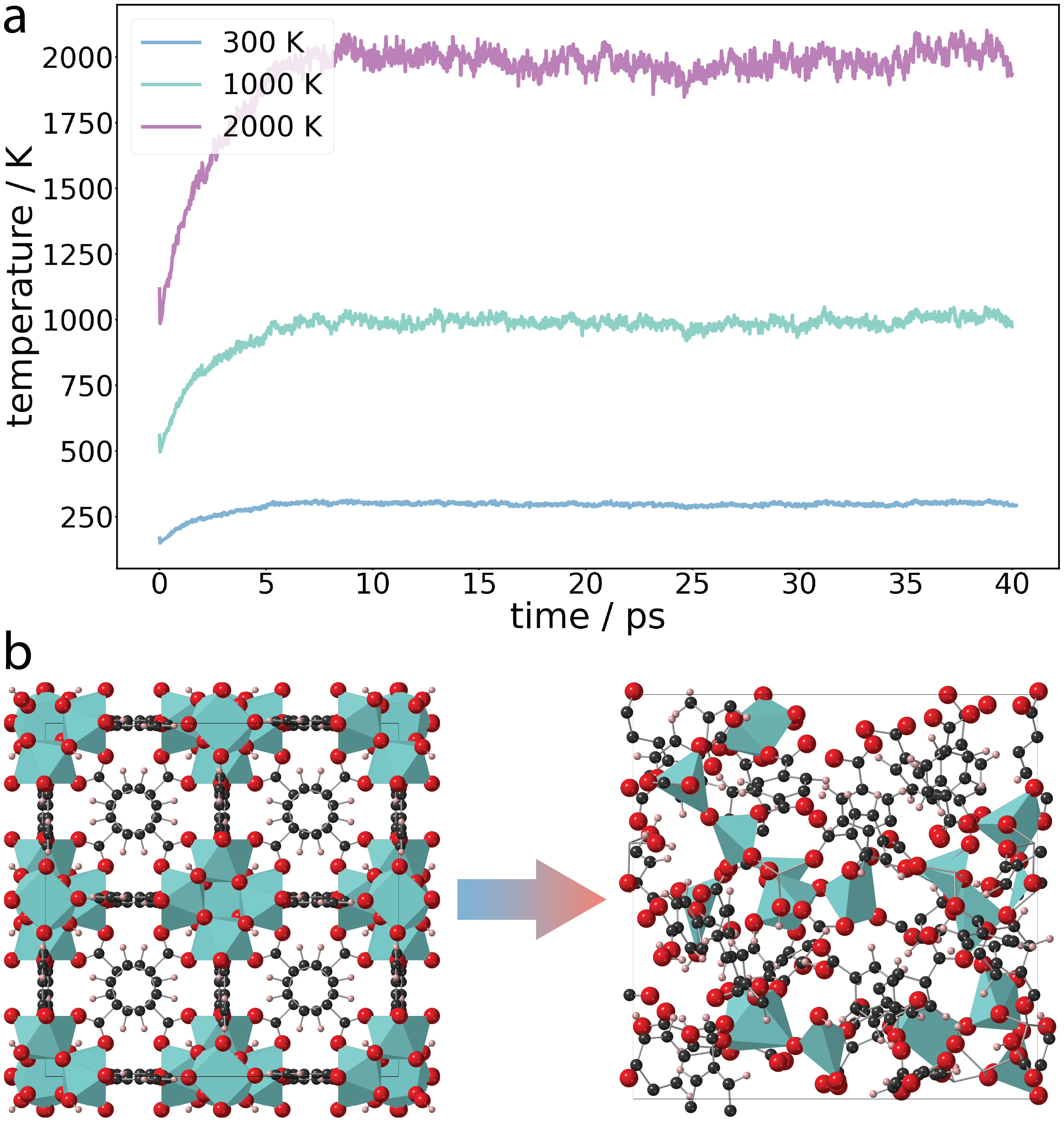}
    \caption{a) Temperature profiles for 40~ps AIMD simulations at 300, 1000 and 2000~K. b) Initial and final structures of UiO-66 taken from a 40~ps AIMD simulation at 2000~K. Zirconium is teal, carbon is black, oxygen is red and hydrogen is pink.}
    \label{fig:temp_decomp}
    \vspace{5mm}
\end{figure}

The dataset presented here contains nine MOFs simulated at 300, 1000 and 2000~K over a relatively long AIMD trajectory of 40~ps (Figure~\ref{fig:temp_decomp}). All systems remain crystalline at 300~K, exhibiting only minor equilibrium fluctuations. This is confirmed by the radial distribution functions (RDF) of the initial and final states and root mean square displacement (RMSD) of atoms over time. The final 300~K RDF shows a retention of sharp peaks, indicating preservation of long range order and RMSD analysis shows only small oscillations below 1~\AA (see S1 of the Supporting Information).

No decomposition products are observed at 1000~K but signs of instability begin to emerge. RMSD increases compared to 300~K, indicating enhanced thermal motion without structural degradation or formation of free species  (see S1 of the Supporting Information). Similarly, RDFs display a partial loss of long range order without significant peak shifts, indicating that the apparent disorder primarily reflects enhanced vibrational motion rather than bond breakage or changes in metal bonding environments. Although these MOFs are reported to decompose below 650 K, the limited timescales accessible to AIMD are not long enough to observe rare atomistic events, such as bond breaking, at this temperature.\cite{10.1021/cm1022882, 10.1021/acs.jpcc.7b10526, 10.1021/ja8057953, 10.1016/j.matt.2020.10.009}

\begin{figure}[tbp]
    \vspace{5mm}
    \centering
    \includegraphics[]{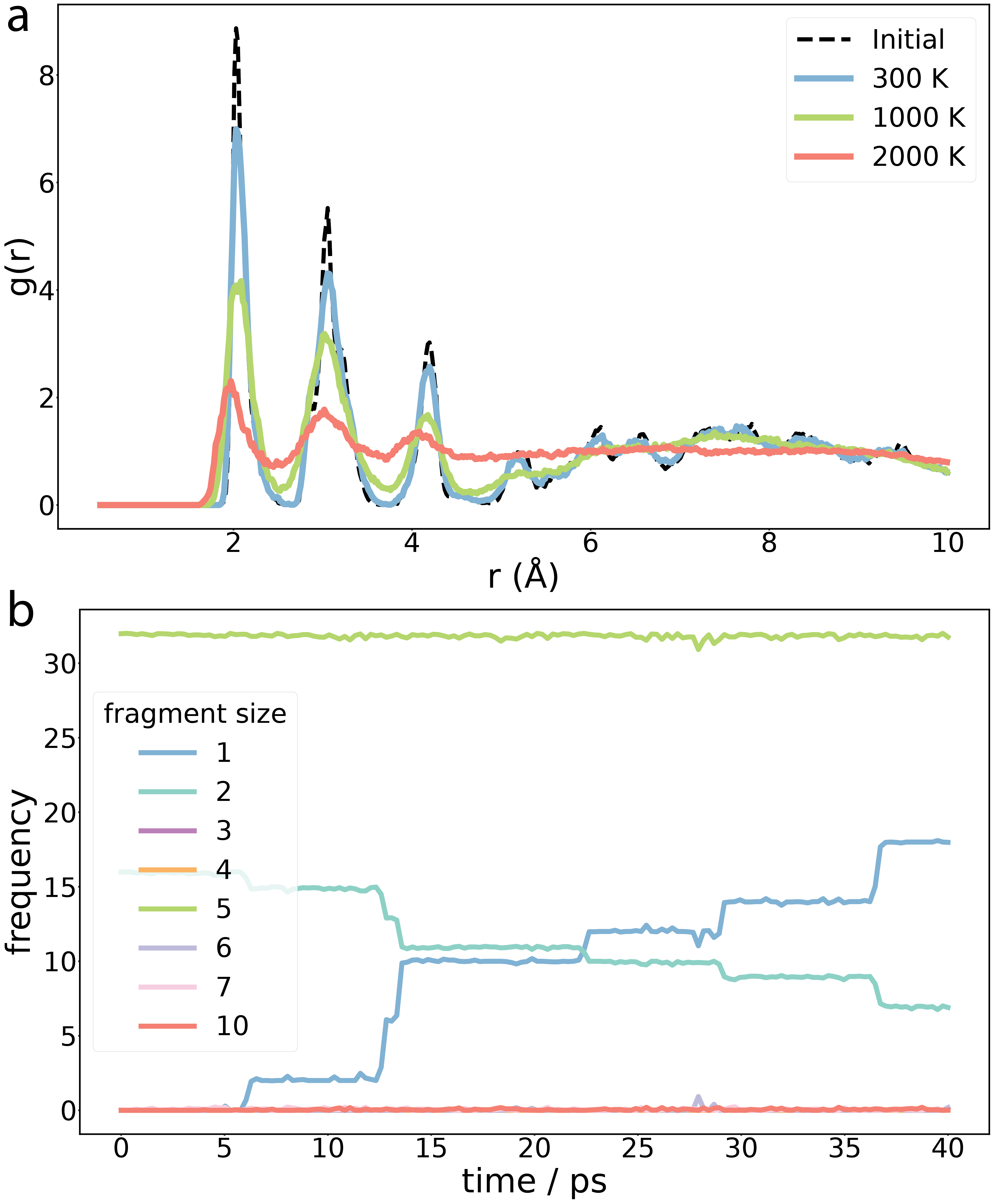}
    \caption{a) RDF of initial crystalline CALF-20 MOF with RDFs from the final structures after 40~ps at 300, 1000 and 2000~K. RDFs are averaged over 50~fs. b) Analysis of CALF-20 linker size over time  showing decomposition of oxalate pillar into \ce{CO2} and the azolate linker persisting during a 40~ps AIMD simulation at 2000~K.}
    \label{fig:struc_analysis}
    \vspace{5mm}
\end{figure}

The 2000~K simulations, in contrast to the lower temperatures, uncover the early stages of MOF decomposition. RDF analysis across all MOFs studied reveals systematic peak shifts corresponding to decreased average metal–ligand bond lengths, alongside the complete loss of long-range order  (Figure~\ref{fig:struc_analysis}, S1 of the Supporting Information). These features confirm the amorphous nature of the resulting structures formed after 40~ps at 2000~K. Additionally, these simulations showed a significant increase in RMSD relative to 300~K, suggesting large scale migration of atoms through the structure. Gas analysis reveals the formation of species such as \ce{CO2}, \ce{CO}, \ce{H2O} and \ce{H2} across CALF-20, MIP-206, UiO-66 and NU-1000  (see S1 of the Supporting Information). Formation of these products indicates that the organic linkers are decomposing at 2000~K on this timescale. Analysis of carbon and nitrogen cluster sizes quantifies the extent of linker decomposition and demonstrates decreases in linker fragment sizes for all MOFs except ZIF-8 (Figure~\ref{fig:struc_analysis}, S1 of the Supporting Information). This indicates that the imidazolate linker present in ZIF-8 has a high thermal stability, as previously reported during 1~ns quench simulations at 1750~K.\cite{10.1002/adts.202500514} Similarly, the azolate linker in CALF-20 remains intact during the 2000~K AIMD simulations, while the oxalate pillar degrades into \ce{CO2}. Beyond linker decomposition, there are also changes in the structure of metal nodes. Across all MOFs, increasing temperature reduces the metal coordination number due to metal–ligand bond cleavage (see S1 of the Supporting Information). 

\subsection{Model accuracy}
We tested five MLIPs on this combined dataset: ORB-v3, MACE-MP-0a, MACE-MPA-0, fairchem ODAC23 and fairchem OMAT. The error was broken down into energy, force and stress components and averaged across all MOFs and temperatures (Table~\ref{tab:model_comparison}). 

\begin{table*}
\caption{Energy, force and stress mean absolute error (MAE) across the investigated universal models averaged over all MOFs and temperatures}
\label{tab:model_comparison}
\resizebox{\textwidth}{!}{
\begin{tabular}{c|c|c|c|c}
Model & energy MAE / meV~atom$^{-1}$ & force MAE / meV\AA$^{-1}$ & stress MAE / MPa & weighted loss\\
\hline
\hline
fairchem OMAT & 3.21 & 94.01 & 293.98 & 11.37\\
ORB-v3 & 3.59 & 119.58 & 325.35 & 14.22\\
fairchem ODAC23 & 11.31 & 188.30 & 401.66 & 25.57\\
MACE-MPA-0 & 12.63 & 200.84 & 336.36 & 27.09\\
MACE-MP-0a & 17.61 & 196.62 & 270.60 & 29.22\\
\end{tabular}
}
\end{table*}

ORB-v3 and fairchem OMAT show low energy mean absolute error (MAE) at 3.59 and 3.21~meV~atom$^{-1}$, respectively, while MACE models and fairchem ODAC23 show errors exceeding 10~meV~atom$^{-1}$. Comparatively, the force and stress MAEs are generally high across all models. ORB-v3 and fairchem OMAT significantly outperform the other uMLIPs, having force MAEs of 119.58 and 94.01~meV\AA$^{-1}$ respectively, which are nearly half the MAE of the other evaluated models. Stress MAE shows less variation between models with values varying around 300~MPa for all models except fairchem ODAC that has an error of 400~MPa. For reference, models finetuned on high temperature MOF datasets have been able to achieve energy MAE of 1~meV~atom$^{-1}$, force MAE below 50 meV~\AA$^{-1}$ and stress MAE between 30 to 300~MPa (dependent on MOF system and not widely reported) across simulations at 1500 and 2000~K.\cite{10.1038/s41524-023-00969-x, 10.1039/d3dd00236e, 10.1038/s41524-024-01427-y,10.1002/adts.202500514, arxiv.2601.16459} Additionally, the loss ranking in Table~\ref{tab:model_comparison} follows model rankings on MATBench, suggesting models tend to have comparable rankings across in- and out-of-domain structures. 

It is important, however, to understand how these models behave across increasing temperatures. Models are generally able to recreate equilibrium dynamics at 300~K. Energy MAEs are below 2 meV atom$^{-1}$ for all models, and stress errors remain comparatively low. However, force MAEs are already high (between 90 and 150 meV \AA$^{-1}$), suggesting that even near equilibrium, force predictions remain substantially less accurate than energy predictions. 

As the temperature increases to 1000~K, errors in energy, force and stress increase across all models. This behavior reflects the emergence of thermally induced motion and structural distortions that are underrepresented in the training datasets of universal models. Differences between models at this temperature become more pronounced. The trends observed in force and stress MAE at 300~K largely persist, with all models exhibiting similar increases in error. By contrast, clearer variation emerges in the energy predictions. The MACE models and fairchem ODAC23 display substantial increases in energy MAE, whereas ORB-v3 and fairchem OMAT maintain errors below 2.5~meV$\,$atom$^{-1}$. As the structural decomposition becomes pronounced, at 2000 K, energy MAE approximately doubles relative to 1000 K, force MAE increases to 150–300 meV \AA$^{-1}$, and stress MAE approaches or exceeds 500 MPa for most models. This behavior strongly indicates that these models are not suitable for simulating early-stage thermal decomposition without additional finetuning.

The relative ranking of models remains largely consistent with temperature. However, fairchem ODAC23 is an exception and outperforms models at 300~K, yet exhibits large error increases at 1000 and 2000~K. Additionally, while MACE-MP-0a slightly outperforms MACE-MPA-0 at lower temperatures, MACE-MPA-0 becomes more accurate at 2000 K, suggesting improved robustness in extreme regimes. 

\begin{figure*}[tbp]
    \vspace{5mm}
    \centering
    \includegraphics[width=0.8\textwidth]{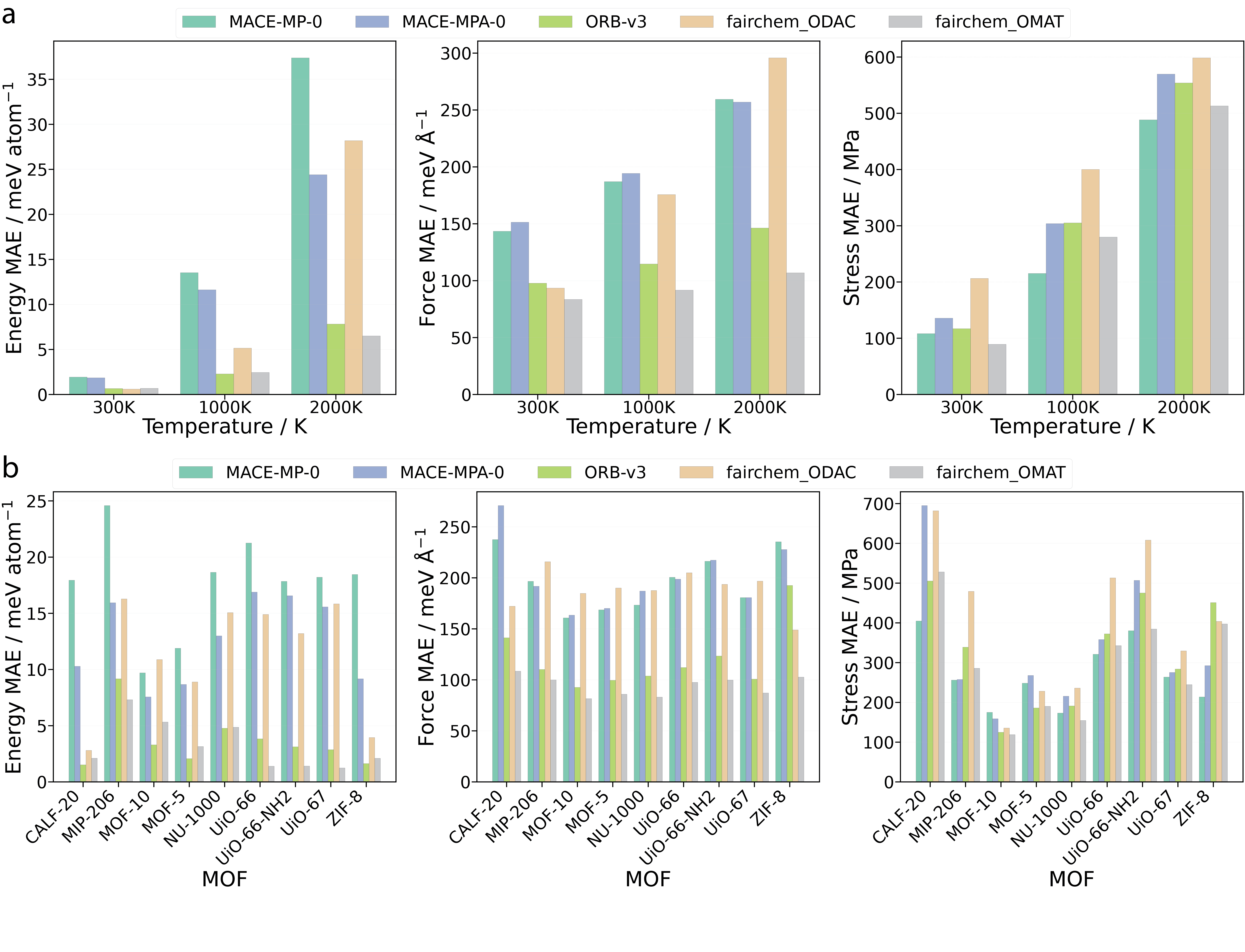}
    \caption{Mean absolute error (MAE) in energy, force, and stress for five uMLIPs benchmarked against nine MOFs at 300, 1000, and 2000 K. a) MAE values averaged across all MOFs at each temperature. b) MAE values averaged across temperatures (300, 1000, and 2000 K) for each MOF.}
    \label{fig:big_mae_fig}
    \vspace{5mm}
\end{figure*}

While model performance improves at lower temperatures, the accuracy is system dependent, with the models exhibiting systematic preference for specific MOFs. For example, CALF-20 and UiO-66 exhibit larger energy and stress errors, whereas MOF-10 and MOF-5 show comparatively lower errors across all models. This correlation reinforces the hypothesis that model accuracy is linked to structural characteristics and environments. Inaccuracies stem from a broader misrepresentation of the underlying potential energy surface rather than prioritizing learning one metric over another (i.e., favoring energy over stress). In contrast, force MAE is relatively constant across MOFs for each model, suggesting a consistent level of error in reproducing interatomic forces irrespective of framework composition or topology. 

Furthermore, all models predict stress with a similar magnitude of error for a given MOF. For instance, stress MAE values for MOF-10 consistently fall between 120 and 150~MPa across models. An interesting consequence of this is that models not trained on stress (fairchem) are able to reproduce stress tensors at the same level of error as models trained on stress. This suggests that models are able to learn stress from energy-force relationships present in the foundational datasets.

Notably, there appears to be no correlation between metal identities and error. Both zinc and zirconium MOFs exhibit comparable error ranges across energy, force and stress MAE. Model performance is likely to be linked to overall structural complexity and thermally activated distortions, rather than the specific metal center. 

\section{Error during a long timescale simulation}
\begin{figure}[tbph]
    \vspace{5mm}
    \centering
    \includegraphics[]{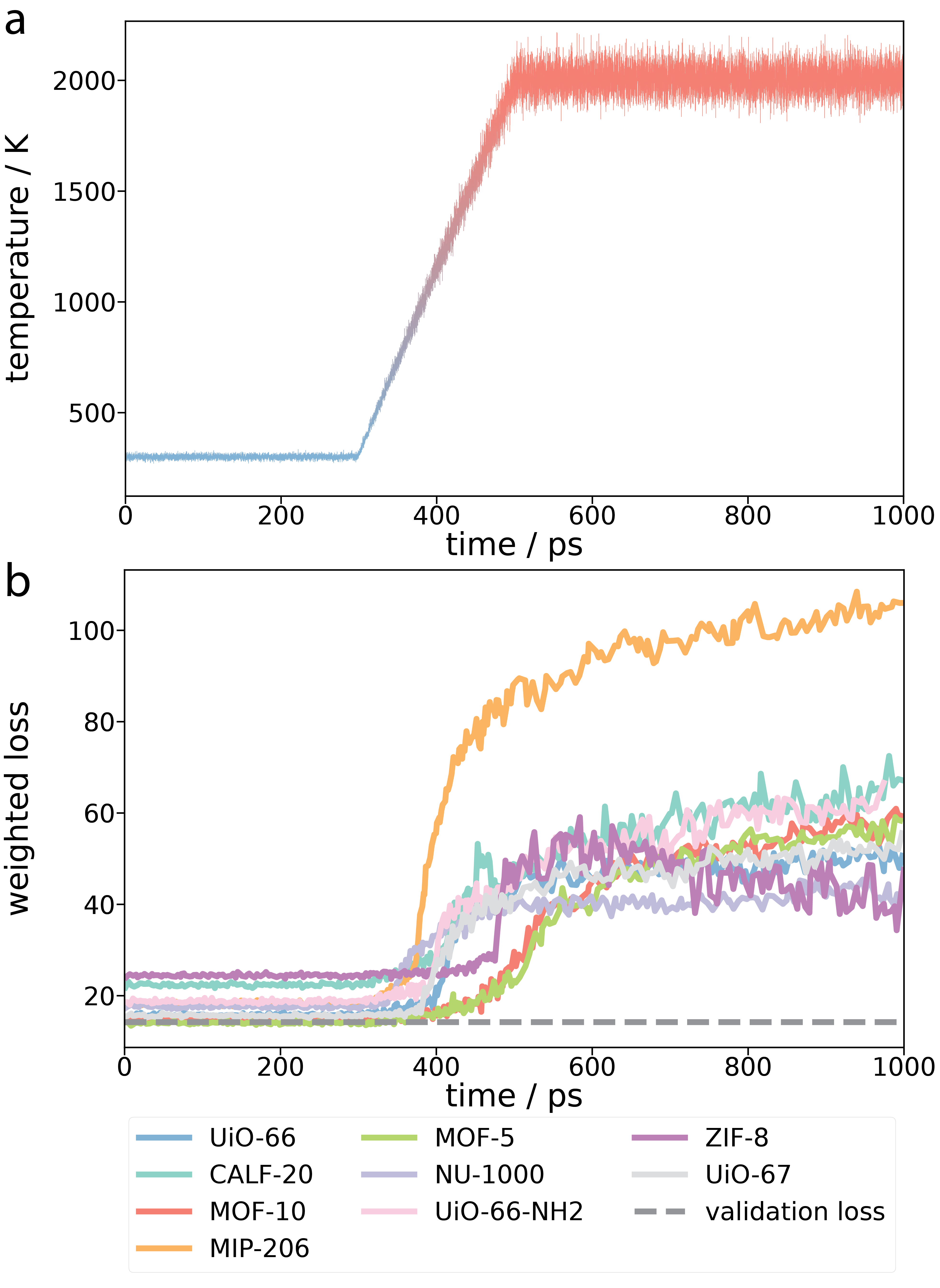}
    \caption{a) Temperature profile during a 1~ns simulation produced with ORB-v3. b) Loss during the 1~ns simulation, showing agreement with AIMD validation tests at low temperature but higher loss as temperature increases.}
    \label{fig:ramp_loss}
    \vspace{5mm}
\end{figure}

To understand whether the errors reported during validation reflect the generative error of these models during MD simulations, ORB-v3 was used to run a series of 1~ns ramp simulations from 300 to 2000~K for each MOF (Figure~\ref{fig:ramp_loss}). Each trajectory consisted of three regimes: equilibration at 300 K (300~ps), a linear temperature ramp from 300 to 2000~K (200~ps), and a 2000 K run (500 ps). From each regime (300 K, ramp, and 2000 K) 100 structures were sampled and evaluated using DFT single-point calculations to construct a reference dataset for model validation.

These simulations demonstrate that ORB-v3 exhibits substantially higher generative error than suggested by static validation metrics using the AIMD data. The energy and stress errors at 300~K are comparable to those obtained from validation, however, the force MAE is noticeably larger. This indicates that even in a stable, low temperature regime where energetics are understood, the dynamics modeled are inaccurate. As a result, the loss during the 300~K MD regime exceeds the magnitude of the ORB-v3 AIMD validation loss (Figure~\ref{fig:ramp_loss} and Table~\ref{tab:model_comparison}). This suggests that structure generation during MD simulations may compound model inaccuracies beyond what is evident from model validation alone.

Nevertheless, the loss at 300~K remains constant over time. As temperature increases the loss also begins to increase, a pronounced jump in error corresponds to the onset of bond breaking, with loss increasing linearly as metal coordination number decreases (Figure~\ref{fig:ramp_loss}, see S2.1 of the Supporting Information). As the temperature reaches and stabilizes at 2000~K the loss does not plateau. Rather, it continues increasing, implying that the amorphous states formed at high temperature are represented with increasing inaccuracy as bonds continue to break (see Section S2.1 of the Supporting Information). Additionally, the weighted loss is three- to four-fold greater than the AIMD validation loss, further exemplifying that these models are ill suited for describing high temperature dynamics (Table~\ref{tab:model_comparison}, Figure~\ref{fig:ramp_loss}). This error increase at 2000~K largely arises from an increase in energy MAE rather than force MAE (see Section S2.2 of the Supporting Information). This trend is consistent with validation results, where the force MAE approximately doubles between 300 and 2000~K, while the energy MAE increases by nearly a factor of 30.

The key finding is that universal models generate trajectories with significantly higher error than implied by the DFT validation. In contrast, finetuned models trained explicitly on 2000~K temperature data have been shown to generate trajectories with error comparable to their validation metrics, highlighting the importance of thermodynamic regime specific training for accurate high temperature dynamics.\cite{arxiv.2601.16459}

\section{Conclusion and outlook}
In this work, we presented a high temperature AIMD dataset spanning nine chemically diverse MOFs at 300, 1000, and 2000~K. This dataset enables systematic benchmarking of universal MLIPs under extreme conditions. Our analysis demonstrates that, while universal models reproduce energies across MOFs, their ability to capture forces and stress tensors deteriorates significantly at elevated temperatures, particularly during structural decomposition and bond breaking events. The discrepancies are evident not only in short AIMD simulations but also during 1~ns ramp simulations, where generative loss continuously increases with time as the structure decomposes.

Among the models tested ORB-v3 and fairchem~{OMAT} perform best near equilibrium, achieving energy MAEs of 0.66 and 0.67 meV atom$^{-1}$, force MAEs of 97.77 and 83.49 meV~\AA$^{-1}$, and stress MAEs of 117.00 and 89.12 MPa, respectively. However, none of the universal models maintain accuracy in high-temperature, non-equilibrium regimes. Importantly, model performance appears largely independent of the metal center (Zn vs. Zr), and instead correlates with the overall structural complexity and thermally activated distortions. Moreover, models not explicitly trained on stress (e.g., fairchem~{ODAC23}) can nevertheless reproduce stress tensors at comparable levels to stress-trained models, highlighting implicit learning of stress from energy-force relationships in foundational datasets. 

Additionally, the use of uMLIP models for molecular dynamics simulations results in substantially higher errors than suggested by validation against reference AIMD datasets. The primary source of discrepancy at lower temperatures arises from inaccurately described interatomic forces, which compromises the accuracy of the dynamics. However, at elevated temperatures, both the energy and force predictions deteriorate leading to progressively larger errors as structural rearrangements and bond-breaking events occur. These results underscore the limitations of current uMLIPs for high-temperature MOF chemistry. While ORB-v3 and fairchem OMAT are suitable for equilibrium energetics, these models struggle to describe early stage decomposition events that are critical for understanding MOF pyrolysis and the formation of MOF-derived catalysts. 

Overall, this paper provides a benchmark for the community and highlights the gap between current uMLIPs and the demands of high-temperature materials modeling. Addressing these limitations will be crucial for deploying uMLIPs as reliable tools for investigating thermally driven transformations in MOFs and other complex porous materials. Going forward, we expect that including high temperature, non-equilibrium and amorphous states in uMLIP training datasets will provide the training coverage necessary to produce more accurate and realistic pyrolysis pathways. 

\section*{Acknowledgements}
J.D.E. is the recipient of an Australian Research Council Discovery Early Career Award (project number DE220100163) funded by the Australian Government. Phoenix HPC service at Adelaide University is thanked for providing high-performance computing resources. This research was supported by the Australian Government's National Collaborative Research Infrastructure Strategy (NCRIS), with access to computational resources provided by Pawsey Supercomputing Research Centre through the National Computational Merit Allocation Scheme. C.W.E is supported by an Australian Government Research Training Program (RTP) Scholarship.

\section*{Supporting Information}
    All code and data for this research are available on Figshare: \href{https://doi.org/10.25909/32061111}{DOI:10.25909/32061111}.  Additional plots and error analysis are provided in the Supporting Information.

\bibliography{references.bib}

@article{10.1021/acs.jpcc.7b10526,
  title = {Influence of Metal–Organic Framework Porosity on Hydrogen Generation from Nanoconfined Ammonia Borane},
  volume = {121},
  ISSN = {1932-7455},
  url = {http://dx.doi.org/10.1021/acs.jpcc.7b10526},
  DOI = {10.1021/acs.jpcc.7b10526},
  number = {49},
  journal = {The Journal of Physical Chemistry C},
  publisher = {American Chemical Society (ACS)},
  author = {Chung,  Jing-Yang and Liao,  Chi-Wei and Chang,  Yi-Wei and Chang,  Bor Kae and Wang,  Hao and Li,  Jing and Wang,  Cheng-Yu},
  year = {2017},
  month = Dec,
  pages = {27369–27378}
}

@misc{10.5281/zenodo.15587498,
author = {Muhammed Shuaibi and Abhishek Das and Anuroop Sriram and Misko and Luis Barroso-Luque and Ray Gao and Siddharth Goyal and Zachary Ulissi and Brandon Wood and Tian Xie and Junwoong Yoon and Brook Wander and Adeesh Kolluru and Richard Barnes and Ethan Sunshine and Kevin Tran and Xiang and Daniel Levine and Nima Shoghi and Ilias Chair and  and Janice Lan and Kaylee Tian and Joseph Musielewicz and clz55 and Weihua Hu and  and Kyle Michel and willis and vbttchr},
doi = {10.5281/zenodo.15587498},
license = {["MIT"]},
title = {{FAIRChem}},
url = {https://github.com/facebookresearch/fairchem},
version = {2.2.0}
}

@article{10.1126/science.1230444,
  title = {The Chemistry and Applications of Metal-Organic Frameworks},
  volume = {341},
  ISSN = {1095-9203},
  url = {http://dx.doi.org/10.1126/science.1230444},
  DOI = {10.1126/science.1230444},
  number = {6149},
  journal = {Science},
  publisher = {American Association for the Advancement of Science (AAAS)},
  author = {Furukawa,  Hiroyasu and Cordova,  Kyle E. and O’Keeffe,  Michael and Yaghi,  Omar M.},
  year = {2013},
  month = aug 
}

@article{10.1021/acs.chemrev.9b00685,
  title = {Metal–Organic Frameworks in Heterogeneous Catalysis: Recent Progress,  New Trends,  and Future Perspectives},
  volume = {120},
  ISSN = {1520-6890},
  url = {http://dx.doi.org/10.1021/acs.chemrev.9b00685},
  DOI = {10.1021/acs.chemrev.9b00685},
  number = {16},
  journal = {Chemical Reviews},
  publisher = {American Chemical Society (ACS)},
  author = {Bavykina,  Anastasiya and Kolobov,  Nikita and Khan,  Il Son and Bau,  Jeremy A. and Ramirez,  Adrian and Gascon,  Jorge},
  year = {2020},
  month = mar,
  pages = {8468–8535}
}

@article{10.1021/acs.jpcc.2c01091,
  title = {Atomistic Models of Amorphous Metal–Organic Frameworks},
  volume = {126},
  ISSN = {1932-7455},
  url = {http://dx.doi.org/10.1021/acs.jpcc.2c01091},
  DOI = {10.1021/acs.jpcc.2c01091},
  number = {16},
  journal = {The Journal of Physical Chemistry C},
  publisher = {American Chemical Society (ACS)},
  author = {Castel,  Nicolas and Coudert,  Fran\c{c}ois-Xavier},
  year = {2022},
  month = apr,
  pages = {6905–6914}
}

@article{10.1039/d3dd00236e,
  title = {Machine learning interatomic potentials for amorphous zeolitic imidazolate frameworks},
  volume = {3},
  ISSN = {2635-098X},
  url = {http://dx.doi.org/10.1039/d3dd00236e},
  DOI = {10.1039/d3dd00236e},
  number = {2},
  journal = {Digital Discovery},
  publisher = {Royal Society of Chemistry (RSC)},
  author = {Castel,  Nicolas and André,  Dune and Edwards,  Connor and Evans,  Jack D. and Coudert,  Fran\c{c}ois-Xavier},
  year = {2024},
  pages = {355–368}
}

@article{10.26434/chemrxiv-2024-n2vzq,
  title = {Towards a Generalizable Machine-Learned Potential for Metal-Organic Frameworks},
  url = {http://dx.doi.org/10.26434/chemrxiv-2024-n2vzq},
  DOI = {10.26434/chemrxiv-2024-n2vzq},
  publisher = {American Chemical Society (ACS)},
  author = {Yue,  Yifei and Saad Aldin,  Mohamed and Loh,  N. Duane and Jiang,  Jianwen},
  year = {2024},
  month = aug 
}

@article{10.1021/acs.jpclett.4c00746,
  title = {Tell Machine Learning Potentials What They Are Needed For: Simulation-Oriented Training Exemplified for Glycine},
  volume = {15},
  ISSN = {1948-7185},
  url = {http://dx.doi.org/10.1021/acs.jpclett.4c00746},
  DOI = {10.1021/acs.jpclett.4c00746},
  number = {16},
  journal = {The Journal of Physical Chemistry Letters},
  publisher = {American Chemical Society (ACS)},
  author = {Ge,  Fuchun and Wang,  Ran and Qu,  Chen and Zheng,  Peikun and Nandi,  Apurba and Conte,  Riccardo and Houston,  Paul L. and Bowman,  Joel M. and Dral,  Pavlo O.},
  year = {2024},
  month = apr,
  pages = {4451–4460}
}

@misc{10.48550/arXiv.2401.00096,
  doi = {10.48550/ARXIV.2401.00096},
  url = {https://arxiv.org/abs/2401.00096},
  author = {Batatia,  Ilyes and Benner,  Philipp and Chiang,  Yuan and Elena,  Alin M. and Kovács,  Dávid P. and Riebesell,  Janosh and Advincula,  Xavier R. and Asta,  Mark and Avaylon,  Matthew and Baldwin,  William J. and Berger,  Fabian and Bernstein,  Noam and Bhowmik,  Arghya and Blau,  Samuel M. and Cărare,  Vlad and Darby,  James P. and De,  Sandip and Della Pia,  Flaviano and Deringer,  Volker L. and Elijošius,  Rokas and El-Machachi,  Zakariya and Falcioni,  Fabio and Fako,  Edvin and Ferrari,  Andrea C. and Genreith-Schriever,  Annalena and George,  Janine and Goodall,  Rhys E. A. and Grey,  Clare P. and Grigorev,  Petr and Han,  Shuang and Handley,  Will and Heenen,  Hendrik H. and Hermansson,  Kersti and Holm,  Christian and Jaafar,  Jad and Hofmann,  Stephan and Jakob,  Konstantin S. and Jung,  Hyunwook and Kapil,  Venkat and Kaplan,  Aaron D. and Karimitari,  Nima and Kermode,  James R. and Kroupa,  Namu and Kullgren,  Jolla and Kuner,  Matthew C. and Kuryla,  Domantas and Liepuoniute,  Guoda and Margraf,  Johannes T. and Magdău,  Ioan-Bogdan and Michaelides,  Angelos and Moore,  J. Harry and Naik,  Aakash A. and Niblett,  Samuel P. and Norwood,  Sam Walton and O'Neill,  Niamh and Ortner,  Christoph and Persson,  Kristin A. and Reuter,  Karsten and Rosen,  Andrew S. and Schaaf,  Lars L. and Schran,  Christoph and Shi,  Benjamin X. and Sivonxay,  Eric and Stenczel,  Tamás K. and Svahn,  Viktor and Sutton,  Christopher and Swinburne,  Thomas D. and Tilly,  Jules and van der Oord,  Cas and Varga-Umbrich,  Eszter and Vegge,  Tejs and Vondrák,  Martin and Wang,  Yangshuai and Witt,  William C. and Zills,  Fabian and Csányi,  Gábor},
  title = {A foundation model for atomistic materials chemistry},
  publisher = {arXiv},
  year = {2024},
  copyright = {Creative Commons Attribution Non Commercial No Derivatives 4.0 International}
}

@misc{10.48550/arxiv.2410.22570,
  doi = {10.48550/ARXIV.2410.22570},
  url = {https://arxiv.org/abs/2410.22570},
  author = {Neumann,  Mark and Gin,  James and Rhodes,  Benjamin and Bennett,  Steven and Li,  Zhiyi and Choubisa,  Hitarth and Hussey,  Arthur and Godwin,  Jonathan},
  title = {Orb: A Fast,  Scalable Neural Network Potential},
  publisher = {arXiv},
  year = {2024},
  copyright = {Creative Commons Attribution 4.0 International}
}

@misc{10.48550/arxiv.2308.14920,
  doi = {10.48550/ARXIV.2308.14920},
  url = {https://arxiv.org/abs/2308.14920},
  author = {Riebesell,  Janosh and Goodall,  Rhys E. A. and Benner,  Philipp and Chiang,  Yuan and Deng,  Bowen and Ceder,  Gerbrand and Asta,  Mark and Lee,  Alpha A. and Jain,  Anubhav and Persson,  Kristin A.},
  title = {Matbench Discovery -- A framework to evaluate machine learning crystal stability predictions},
  publisher = {arXiv},
  year = {2023},
  copyright = {Creative Commons Attribution 4.0 International}
}

@article{10.1016/j.cpc.2004.12.014,
  title = {Quickstep: Fast and accurate density functional calculations using a mixed Gaussian and plane waves approach},
  volume = {167},
  ISSN = {0010-4655},
  url = {http://dx.doi.org/10.1016/j.cpc.2004.12.014},
  DOI = {10.1016/j.cpc.2004.12.014},
  number = {2},
  journal = {Computer Physics Communications},
  publisher = {Elsevier BV},
  author = {VandeVondele,  Joost and Krack,  Matthias and Mohamed,  Fawzi and Parrinello,  Michele and Chassaing,  Thomas and Hutter,  J\"{u}rg},
  year = {2005},
  month = apr,
  pages = {103–128}
}

@article{10.1063/5.0007045,
  title = {CP2K: An electronic structure and molecular dynamics software package - Quickstep: Efficient and accurate electronic structure calculations},
  volume = {152},
  ISSN = {1089-7690},
  url = {http://dx.doi.org/10.1063/5.0007045},
  DOI = {10.1063/5.0007045},
  number = {19},
  journal = {The Journal of Chemical Physics},
  publisher = {AIP Publishing},
  author = {K\"{u}hne,  Thomas D. and Iannuzzi,  Marcella and Del Ben,  Mauro and Rybkin,  Vladimir V. and Seewald,  Patrick and Stein,  Frederick and Laino,  Teodoro and Khaliullin,  Rustam Z. and Sch\"{u}tt,  Ole and Schiffmann,  Florian and Golze,  Dorothea and Wilhelm,  Jan and Chulkov,  Sergey and Bani-Hashemian,  Mohammad Hossein and Weber,  Valéry and Borštnik,  Urban and Taillefumier,  Mathieu and Jakobovits,  Alice Shoshana and Lazzaro,  Alfio and Pabst,  Hans and M\"{u}ller,  Tiziano and Schade,  Robert and Guidon,  Manuel and Andermatt,  Samuel and Holmberg,  Nico and Schenter,  Gregory K. and Hehn,  Anna and Bussy,  Augustin and Belleflamme,  Fabian and Tabacchi,  Gloria and Gl\"{o}ß,  Andreas and Lass,  Michael and Bethune,  Iain and Mundy,  Christopher J. and Plessl,  Christian and Watkins,  Matt and VandeVondele,  Joost and Krack,  Matthias and Hutter,  J\"{u}rg},
  year = {2020},
  month = may 
}

@article{10.1103/physrevb.54.1703,
  title = {Separable dual-space Gaussian pseudopotentials},
  volume = {54},
  ISSN = {1095-3795},
  url = {http://dx.doi.org/10.1103/PhysRevB.54.1703},
  DOI = {10.1103/physrevb.54.1703},
  number = {3},
  journal = {Physical Review B},
  publisher = {American Physical Society (APS)},
  author = {Goedecker,  S. and Teter,  M. and Hutter,  J.},
  year = {1996},
  month = jul,
  pages = {1703–1710}
}

@article{10.1103/physrevlett.77.3865,
  title = {Generalized Gradient Approximation Made Simple},
  volume = {77},
  ISSN = {1079-7114},
  url = {http://dx.doi.org/10.1103/PhysRevLett.77.3865},
  DOI = {10.1103/physrevlett.77.3865},
  number = {18},
  journal = {Physical Review Letters},
  publisher = {American Physical Society (APS)},
  author = {Perdew,  John P. and Burke,  Kieron and Ernzerhof,  Matthias},
  year = {1996},
  month = oct,
  pages = {3865–3868}
}

@article{10.1063/1.3382344,
  title = {A consistent and accurateab initioparametrization of density functional dispersion correction (DFT-D) for the 94 elements H-Pu},
  volume = {132},
  ISSN = {1089-7690},
  url = {http://dx.doi.org/10.1063/1.3382344},
  DOI = {10.1063/1.3382344},
  number = {15},
  journal = {The Journal of Chemical Physics},
  publisher = {AIP Publishing},
  author = {Grimme,  Stefan and Antony,  Jens and Ehrlich,  Stephan and Krieg,  Helge},
  year = {2010},
  month = apr 
}

@article{10.1063/1.2408420,
  title = {Canonical sampling through velocity rescaling},
  volume = {126},
  ISSN = {1089-7690},
  url = {http://dx.doi.org/10.1063/1.2408420},
  DOI = {10.1063/1.2408420},
  number = {1},
  journal = {The Journal of Chemical Physics},
  publisher = {AIP Publishing},
  author = {Bussi,  Giovanni and Donadio,  Davide and Parrinello,  Michele},
  year = {2007},
  month = jan 
}

@article{10.1038/s41524-023-00969-x,
  title = {Machine learning potentials for metal-organic frameworks using an incremental learning approach},
  volume = {9},
  ISSN = {2057-3960},
  url = {http://dx.doi.org/10.1038/s41524-023-00969-x},
  DOI = {10.1038/s41524-023-00969-x},
  number = {1},
  journal = {npj Computational Materials},
  publisher = {Springer Science and Business Media LLC},
  author = {Vandenhaute,  Sander and Cools-Ceuppens,  Maarten and DeKeyser,  Simon and Verstraelen,  Toon and Van Speybroeck,  Veronique},
  year = {2023},
  month = feb 
}

@article{ase-paper,
  author={Ask Hjorth Larsen and Jens Jørgen Mortensen and Jakob Blomqvist and Ivano E Castelli and Rune Christensen and Marcin
Dułak and Jesper Friis and Michael N Groves and Bjørk Hammer and Cory Hargus and Eric D Hermes and Paul C Jennings and Peter
Bjerre Jensen and James Kermode and John R Kitchin and Esben Leonhard Kolsbjerg and Joseph Kubal and Kristen
Kaasbjerg and Steen Lysgaard and Jón Bergmann Maronsson and Tristan Maxson and Thomas Olsen and Lars Pastewka and Andrew
Peterson and Carsten Rostgaard and Jakob Schiøtz and Ole Schütt and Mikkel Strange and Kristian S Thygesen and Tejs
Vegge and Lasse Vilhelmsen and Michael Walter and Zhenhua Zeng and Karsten W Jacobsen},
  title={The atomic simulation environment—a Python library for working with atoms},
  journal={Journal of Physics: Condensed Matter},
  volume={29},
  number={27},
  pages={273002},
  url={http://stacks.iop.org/0953-8984/29/i=27/a=273002},
  year={2017}
}

@article{10.1038/s41524-024-01427-y,
  title = {Quantum-accurate machine learning potentials for metal-organic frameworks using temperature driven active learning},
  volume = {10},
  ISSN = {2057-3960},
  url = {http://dx.doi.org/10.1038/s41524-024-01427-y},
  DOI = {10.1038/s41524-024-01427-y},
  number = {1},
  journal = {npj Computational Materials},
  publisher = {Springer Science and Business Media LLC},
  author = {Sharma,  Abhishek and Sanvito,  Stefano},
  year = {2024},
  month = oct 
}

@article{10.1016/j.ccr.2020.213319,
  title = {Metal-organic framework (MOF)-derived catalysts for fine chemical production},
  volume = {416},
  ISSN = {0010-8545},
  url = {http://dx.doi.org/10.1016/j.ccr.2020.213319},
  DOI = {10.1016/j.ccr.2020.213319},
  journal = {Coordination Chemistry Reviews},
  publisher = {Elsevier BV},
  author = {Konnerth,  Hannelore and Matsagar,  Babasaheb M. and Chen,  Season S. and Prechtl,  Martin H.G. and Shieh,  Fa-Kuen and Wu,  Kevin C.-W.},
  year = {2020},
  month = aug,
  pages = {213319}
}

@article{10.1021/ar5000314,
  title = {Amorphous Metal–Organic Frameworks},
  volume = {47},
  ISSN = {1520-4898},
  url = {http://dx.doi.org/10.1021/ar5000314},
  DOI = {10.1021/ar5000314},
  number = {5},
  journal = {Accounts of Chemical Research},
  publisher = {American Chemical Society (ACS)},
  author = {Bennett,  Thomas D. and Cheetham,  Anthony K.},
  year = {2014},
  month = apr,
  pages = {1555–1562}
}

@article{10.1039/C7CP08508G,
  title = {Structural investigations of amorphous metal–organic frameworks formed via different routes},
  volume = {20},
  ISSN = {1463-9084},
  url = {http://dx.doi.org/10.1039/C7CP08508G},
  DOI = {10.1039/c7cp08508g},
  number = {11},
  journal = {Physical Chemistry Chemical Physics},
  publisher = {Royal Society of Chemistry (RSC)},
  author = {Keen,  D. A. and Bennett,  T. D.},
  year = {2018},
  pages = {7857–7861}
}

@article{10.1038/s41563-020-0777-6,
  title = {Machine-learned potentials for next-generation matter simulations},
  volume = {20},
  ISSN = {1476-4660},
  url = {http://dx.doi.org/10.1038/s41563-020-0777-6},
  DOI = {10.1038/s41563-020-0777-6},
  number = {6},
  journal = {Nature Materials},
  publisher = {Springer Science and Business Media LLC},
  author = {Friederich,  Pascal and H\"{a}se,  Florian and Proppe,  Jonny and Aspuru-Guzik,  Alán},
  year = {2021},
  month = may,
  pages = {750–761}
}

@article{10.1021/jacs.9b03234,
  title = {Rich Polymorphism of a Metal–Organic Framework in Pressure–Temperature Space},
  volume = {141},
  ISSN = {1520-5126},
  url = {http://dx.doi.org/10.1021/jacs.9b03234},
  DOI = {10.1021/jacs.9b03234},
  number = {23},
  journal = {Journal of the American Chemical Society},
  publisher = {American Chemical Society (ACS)},
  author = {Widmer,  Remo N. and Lampronti,  Giulio I. and Chibani,  Siwar and Wilson,  Craig W. and Anzellini,  Simone and Farsang,  Stefan and Kleppe,  Annette K. and Casati,  Nicola P. M. and MacLeod,  Simon G. and Redfern,  Simon A. T. and Coudert,  Fran\c{c}ois-Xavier and Bennett,  Thomas D.},
  year = {2019},
  month = may,
  pages = {9330–9337}
}

@article{10.1021/acsami.5c10085,
  title = {Understanding the Role of the Zr-MOF Support Structure on Templated Ternary CO2 Hydrogenation Catalyst Structure and Activity},
  ISSN = {1944-8252},
  url = {http://dx.doi.org/10.1021/acsami.5c10085},
  DOI = {10.1021/acsami.5c10085},
  journal = {ACS Applied Materials \& Interfaces},
  publisher = {American Chemical Society (ACS)},
  author = {Linder-Patton,  Oliver M. and Wang,  Lizhuo and Evans,  Jack D. and Yasin,  Nor Hafizah and Berahim-Jusoh,  Nor Hafizah and Li,  Siqi and Huang,  Jun and Phak,  Chan Zhe and Seman,  Akbar A. and Sumby,  Christopher J. and Doonan,  Christian J.},
  year = {2025},
  month = jul 
}

@article{10.1002/adts.202500514,
  title = {Exploring Foundational Machine Learned Potentials for Treating the High Temperature Dynamics of Metal‐Organic Frameworks},
  ISSN = {2513-0390},
  url = {http://dx.doi.org/10.1002/adts.202500514},
  DOI = {10.1002/adts.202500514},
  journal = {Advanced Theory and Simulations},
  publisher = {Wiley},
  author = {Edwards,  Connor W. and Evans,  Jack D.},
  year = {2025},
  month = oct 
}

@inproceedings{Batatia2022mace,
  title={{MACE}: Higher Order Equivariant Message Passing Neural Networks for Fast and Accurate Force Fields},
  author={Ilyes Batatia and David Peter Kovacs and Gregor N. C. Simm and Christoph Ortner and Gabor Csanyi},
  booktitle={Advances in Neural Information Processing Systems},
  editor={Alice H. Oh and Alekh Agarwal and Danielle Belgrave and Kyunghyun Cho},
  year={2022},
  url={https://openreview.net/forum?id=YPpSngE-ZU}
}

@misc{10.48550/arXiv.2205.06643,
  title = {The Design Space of E(3)-Equivariant Atom-Centered Interatomic Potentials},
  author = {Batatia, Ilyes and Batzner, Simon and Kov{\'a}cs, D{\'a}vid P{\'e}ter and Musaelian, Albert and Simm, Gregor N. C. and Drautz, Ralf and Ortner, Christoph and Kozinsky, Boris and Cs{\'a}nyi, G{\'a}bor},
  year = {2022},
  number = {arXiv:2205.06643},
  eprint = {2205.06643},
  eprinttype = {arxiv},
  doi = {10.48550/arXiv.2205.06643},
  archiveprefix = {arXiv}
 }

@misc{10.48550/ARXIV.2504.06231,
      title={Orb-v3: atomistic simulation at scale}, 
      author={Benjamin Rhodes and Sander Vandenhaute and Vaidotas Šimkus and James Gin and Jonathan Godwin and Tim Duignan and Mark Neumann},
      year={2025},
      eprint={2504.06231},
      archivePrefix={arXiv},
      primaryClass={cond-mat.mtrl-sci},
      url={https://arxiv.org/abs/2504.06231}, 
      doi = {10.48550/ARXIV.2504.06231},
}

@article{10.1021/jacs.4c14455,
	title = {Data-Efficient Multifidelity Training for High-Fidelity Machine Learning Interatomic Potentials},
	volume = {147},
	doi = {10.1021/jacs.4c14455},
	number = {1},
	journal = {J. Am. Chem. Soc.},
	author = {Kim, Jaesun and Kim, Jisu and Kim, Jaehoon and Lee, Jiho and Park, Yutack and Kang, Youngho and Han, Seungwu},
	year = {2024},
	pages = {1042--1054},
}

@article{10.1038/nature01650,
  title = {Reticular synthesis and the design of new materials},
  volume = {423},
  ISSN = {1476-4687},
  url = {http://dx.doi.org/10.1038/nature01650},
  DOI = {10.1038/nature01650},
  number = {6941},
  journal = {Nature},
  publisher = {Springer Science and Business Media LLC},
  author = {Yaghi,  Omar M. and O’Keeffe,  Michael and Ockwig,  Nathan W. and Chae,  Hee K. and Eddaoudi,  Mohamed and Kim,  Jaheon},
  year = {2003},
  month = jun,
  pages = {705–714}
}

@article{10.1021/ja8057953,
  title = {A New Zirconium Inorganic Building Brick Forming Metal Organic Frameworks with Exceptional Stability},
  volume = {130},
  ISSN = {1520-5126},
  url = {http://dx.doi.org/10.1021/ja8057953},
  DOI = {10.1021/ja8057953},
  number = {42},
  journal = {Journal of the American Chemical Society},
  publisher = {American Chemical Society (ACS)},
  author = {Cavka,  Jasmina Hafizovic and Jakobsen,  Søren and Olsbye,  Unni and Guillou,  Nathalie and Lamberti,  Carlo and Bordiga,  Silvia and Lillerud,  Karl Petter},
  year = {2008},
  month = sep,
  pages = {13850–13851}
}

@article{10.1016/j.matt.2020.10.009,
  title = {A Mesoporous Zirconium-Isophthalate Multifunctional Platform},
  volume = {4},
  ISSN = {2590-2385},
  url = {http://dx.doi.org/10.1016/j.matt.2020.10.009},
  DOI = {10.1016/j.matt.2020.10.009},
  number = {1},
  journal = {Matter},
  publisher = {Elsevier BV},
  author = {Wang,  Sujing and Chen,  Liyu and Wahiduzzaman,  Mohammad and Tissot,  Antoine and Zhou,  Lin and Ibarra,  Ilich A. and Gutiérrez-Alejandre,  Aída and Lee,  Ji Sun and Chang,  Jong-San and Liu,  Zheng and Marrot,  Jér\^ome and Shepard,  William and Maurin,  Guillaume and Xu,  Qiang and Serre,  Christian},
  year = {2021},
  month = jan,
  pages = {182–194}
}

@misc{salex_2024,
  title = {{sAlex: a Matbench-Discovery compliant subsample of the Alexandria dataset}},
  howpublished = {\url{https://matbench-discovery.materialsproject.org/data/salex}},
  year  = {2024},
  note  = {sAlex dataset used in MACE second-generation models. Accessed 2025-10-30}
}

@misc{mptrj_2023,
  title = {{Materials Project Trajectory (MPtrj) dataset}},
  howpublished = {\url{https://figshare.com/articles/dataset/Materials_Project_Trjectory_MPtrj_Dataset/23713842}},
  year  = {2023},
  note  = {Figshare dataset (MPtrj) — accessed 2025-10-30}
}

@article{Konnerth2020,
  title = {Metal-organic framework (MOF)-derived catalysts for fine chemical production},
  volume = {416},
  ISSN = {0010-8545},
  url = {http://dx.doi.org/10.1016/j.ccr.2020.213319},
  DOI = {10.1016/j.ccr.2020.213319},
  journal = {Coordination Chemistry Reviews},
  publisher = {Elsevier BV},
  author = {Konnerth,  Hannelore and Matsagar,  Babasaheb M. and Chen,  Season S. and Prechtl,  Martin H.G. and Shieh,  Fa-Kuen and Wu,  Kevin C.-W.},
  year = {2020},
  month = aug,
  pages = {213319}
}

@article{Li2021,
  title = {Transition metal-based bimetallic MOFs and MOF-derived catalysts for electrochemical oxygen evolution reaction},
  volume = {14},
  ISSN = {1754-5706},
  url = {http://dx.doi.org/10.1039/D0EE03697H},
  DOI = {10.1039/d0ee03697h},
  number = {4},
  journal = {Energy \& Environmental Science},
  publisher = {Royal Society of Chemistry (RSC)},
  author = {Li,  Songsong and Gao,  Yangqin and Li,  Ning and Ge,  Lei and Bu,  Xianhui and Feng,  Pingyun},
  year = {2021},
  pages = {1897–1927}
}

@article{Deng2025,
  title = {Systematic softening in universal machine learning interatomic potentials},
  volume = {11},
  ISSN = {2057-3960},
  url = {http://dx.doi.org/10.1038/s41524-024-01500-6},
  DOI = {10.1038/s41524-024-01500-6},
  number = {1},
  journal = {npj Computational Materials},
  publisher = {Springer Science and Business Media LLC},
  author = {Deng,  Bowen and Choi,  Yunyeong and Zhong,  Peichen and Riebesell,  Janosh and Anand,  Shashwat and Li,  Zhuohan and Jun,  KyuJung and Persson,  Kristin A. and Ceder,  Gerbrand},
  year = {2025},
  month = jan 
}

@misc{arxiv.2601.16459,
  doi = {10.48550/ARXIV.2601.16459},
  url = {https://arxiv.org/abs/2601.16459},
  author = {Edwards,  Connor W. and Linder-Patton,  Oliver M. and Evans,  Jack D.},
  title = {Simulations of High Temperature Decomposition of Metal-Organic Frameworks to form Amorphous Catalysts},
  publisher = {arXiv},
  year = {2026},
  copyright = {Creative Commons Attribution Non Commercial No Derivatives 4.0 International}
}

@article{Kra2025,
  title = {MOFSimBench: evaluating universal machine learning interatomic potentials in metal-organic framework molecular modeling},
  volume = {12},
  ISSN = {2057-3960},
  url = {http://dx.doi.org/10.1038/s41524-025-01872-3},
  DOI = {10.1038/s41524-025-01872-3},
  number = {1},
  journal = {npj Computational Materials},
  publisher = {Springer Science and Business Media LLC},
  author = {Kraß,  Hendrik and Huang,  Ju and Moosavi,  Seyed Mohamad},
  year = {2025},
  month = dec 
}

@article{10.1126/science.abn3445,
  title = {The central role of density functional theory in the AI age},
  volume = {381},
  ISSN = {1095-9203},
  url = {http://dx.doi.org/10.1126/science.abn3445},
  DOI = {10.1126/science.abn3445},
  number = {6654},
  journal = {Science},
  publisher = {American Association for the Advancement of Science (AAAS)},
  author = {Huang,  Bing and von Rudorff,  Guido Falk and von Lilienfeld,  O. Anatole},
  year = {2023},
  month = jul,
  pages = {170–175}
}

@misc{10.48550/arxiv.2602.13725,
  doi = {10.48550/ARXIV.2602.13725},
  url = {https://arxiv.org/abs/2602.13725},
  author = {Edwards,  Connor W. and Yang,  Fengxu and Stracke,  Konstantin and Evans,  Jack D.},
  title = {MLIP-MC: A Framework for Adsorption Simulations using Machine-Learned Interatomic Potentials},
  publisher = {arXiv},
  year = {2026},
  copyright = {Creative Commons Attribution Non Commercial No Derivatives 4.0 International}
}

@article{10.48550/arXiv.2410.12771,
  title   = {Open Materials 2024 (OMat24) Inorganic Materials Dataset and Models},
  author  = {Barroso-Luque, Luis and Shuaibi, Muhammed and Fu, Xiang and Wood, Brandon M. and Dzamba, Misko and Gao, Meng and Rizvi, Ammar and Zitnick, C. Lawrence and Ulissi, Zachary W.},
  date    = {2024-10-16},
  eprint  = {2410.12771},
  eprinttype = {arXiv},
  doi     = {10.48550/arXiv.2410.12771},
  url     = {http://arxiv.org/abs/2410.12771}
}

@misc{10.48550/arxiv.2508.03162,
  doi = {10.48550/ARXIV.2508.03162},
  url = {https://arxiv.org/abs/2508.03162},
  author = {Sriram,  Anuroop and Brabson,  Logan M. and Yu,  Xiaohan and Choi,  Sihoon and Abdelmaqsoud,  Kareem and Moubarak,  Elias and de Haan,  Pim and L\"{o}we,  Sindy and Brehmer,  Johann and Kitchin,  John R. and Welling,  Max and Zitnick,  C. Lawrence and Ulissi,  Zachary and Medford,  Andrew J. and Sholl,  David S.},
  title = {The Open DAC 2025 Dataset for Sorbent Discovery in Direct Air Capture},
  publisher = {arXiv},
  year = {2025},
  copyright = {Creative Commons Attribution 4.0 International}
}

@article{10.1021/cm1022882,
  title = {Disclosing the Complex Structure of UiO-66 Metal Organic Framework: A Synergic Combination of Experiment and Theory},
  volume = {23},
  ISSN = {1520-5002},
  url = {http://dx.doi.org/10.1021/cm1022882},
  DOI = {10.1021/cm1022882},
  number = {7},
  journal = {Chemistry of Materials},
  publisher = {American Chemical Society (ACS)},
  author = {Valenzano,  Loredana and Civalleri,  Bartolomeo and Chavan,  Sachin and Bordiga,  Silvia and Nilsen,  Merete H. and Jakobsen,  Søren and Lillerud,  Karl Petter and Lamberti,  Carlo},
  year = {2011},
  month = mar,
  pages = {1700–1718}
}

@article{10.1021/acscatal.6b01222,
  title = {Development of MOF-Derived Carbon-Based Nanomaterials for Efficient Catalysis},
  volume = {6},
  ISSN = {2155-5435},
  url = {http://dx.doi.org/10.1021/acscatal.6b01222},
  DOI = {10.1021/acscatal.6b01222},
  number = {9},
  journal = {ACS Catalysis},
  publisher = {American Chemical Society (ACS)},
  author = {Shen,  Kui and Chen,  Xiaodong and Chen,  Junying and Li,  Yingwei},
  year = {2016},
  month = aug,
  pages = {5887–5903}
}

@article{Bavykina2020,
  title = {Metal–Organic Frameworks in Heterogeneous Catalysis: Recent Progress,  New Trends,  and Future Perspectives},
  volume = {120},
  ISSN = {1520-6890},
  url = {http://dx.doi.org/10.1021/acs.chemrev.9b00685},
  DOI = {10.1021/acs.chemrev.9b00685},
  number = {16},
  journal = {Chemical Reviews},
  publisher = {American Chemical Society (ACS)},
  author = {Bavykina,  Anastasiya and Kolobov,  Nikita and Khan,  Il Son and Bau,  Jeremy A. and Ramirez,  Adrian and Gascon,  Jorge},
  year = {2020},
  month = mar,
  pages = {8468–8535}
}

@article{Loew2025,
  title = {Universal machine learning interatomic potentials are ready for phonons},
  volume = {11},
  ISSN = {2057-3960},
  url = {http://dx.doi.org/10.1038/s41524-025-01650-1},
  DOI = {10.1038/s41524-025-01650-1},
  number = {1},
  journal = {npj Computational Materials},
  publisher = {Springer Science and Business Media LLC},
  author = {Loew,  Antoine and Sun,  Dewen and Wang,  Hai-Chen and Botti,  Silvana and Marques,  Miguel A. L.},
  year = {2025},
  month = jun 
}

@misc{10.48550/arxiv.2511.22885,
  doi = {10.48550/ARXIV.2511.22885},
  url = {https://arxiv.org/abs/2511.22885},
  author = {Stracke,  Konstantin and Edwards,  Connor W. and Evans,  Jack D.},
  title = {Evaluating Mechanical Property Prediction across Material Classes using Molecular Dynamics Simulations with Universal Machine-Learned Interatomic Potentials},
  publisher = {arXiv},
  year = {2025},
  copyright = {Creative Commons Attribution Non Commercial No Derivatives 4.0 International}
}

\end{document}